\documentclass[conference]{IEEEtran}
\IEEEoverridecommandlockouts

\usepackage[utf8]{inputenc}
\usepackage[T1]{fontenc}
\usepackage{cite}
\usepackage{amsmath,amssymb}
\usepackage{graphicx}
\usepackage{booktabs}
\usepackage{tabularx}
\usepackage{array}
\usepackage{xurl}
\usepackage{xcolor}
\usepackage{microtype}
\usepackage[hidelinks]{hyperref}

\newcolumntype{Y}{>{\raggedright\arraybackslash}X}

\renewcommand{\baselinestretch}{0.9}

\begin{document}

\title{AI-Native Open RAN: A Roadmap from xApps and rApps to Autonomous Network Agents%
\thanks{This work was supported, in part, at Clemson University by the State of South Carolina through funding for the Battelle Savannah River Alliance Workforce Development Program and NVIDIA Academic Development Grant. It is also funded by the National Science Foundation under Grant Numbers CNS-2202972, CNS- 2318726, and CNS-2232048.}}

\author{
  \IEEEauthorblockN{
    Ryan Barker,
    Alireza Ebrahimi Dorcheh,
    Tolunay Seyfi,
    Mohammad Raihan Uddin,\\
    Alireza Mohammadhosseini,
    Julia Boone,
    Stephen Streit,
    Drew Schlesener,
    Fatemeh Afghah
  }
  \IEEEauthorblockA{
    Holcombe Department of Electrical and Computer Engineering,
    Clemson University, Clemson, SC, USA\\
  }
}

\maketitle

\begin{abstract}
Open Radio Access Networks (O-RAN) have emerged as a transformative paradigm for future wireless systems by introducing openness, virtualization, disaggregation, and programmable intelligence through the RAN Intelligent Controller (RIC). The availability of standardized interfaces and near-real-time control loops has created unprecedented opportunities for integrating artificial intelligence (AI) into radio access network management and optimization. Over the past several years, a broad range of AI techniques have been proposed to address key O-RAN challenges such as radio resource management, network slicing, traffic prediction, mobility management, interference mitigation, and spectrum sharing. Despite significant progress, existing solutions often remain task-specific, require extensive retraining, and exhibit limited generalization across deployment environments and network conditions.

This paper presents a comprehensive review of AI-enabled O-RAN systems and provides a unifying perspective on the evolution of intelligence in wireless networks. We first examine the O-RAN architecture and the role of intelligence within near-real-time and non-real-time RIC frameworks. We then develop a taxonomy of AI approaches for O-RAN, covering machine learning, deep reinforcement learning (DRL), digital-twin-assisted optimization, and emerging foundation-model-based architectures.
\end{abstract}

\begin{IEEEkeywords}
AI-native RAN (AI-RAN), Open RAN (O-RAN), RAN Intelligent Controller (RIC), xApp, rApp, dApp, reinforcement learning.
\end{IEEEkeywords}





\section{Introduction}

Open Radio Access Network (Open RAN) changes the radio access network from a tightly integrated appliance into a programmable system assembled from disaggregated functions, open interfaces, cloud platforms, and independently developed control applications. The O-RAN Alliance architecture makes this programmability operational through the service management and orchestration framework, the non-real-time RAN Intelligent Controller (non-RT RIC), the near-real-time RIC (near-RT RIC), and E2-connected radio nodes. These components create distinct places to analyze long-horizon data, train and govern models, and execute closed-loop actions. Surveys of the architecture have documented the resulting interfaces, algorithms, security questions, and deployment opportunities, while broader 6G analyses identify Open RAN as a practical substrate for increasingly software-defined cellular systems~\cite{polese2023understanding,polese2024empowering}.

The central research opportunity is not simply to place artificial intelligence (AI) inside the RIC. It is to determine which intelligence belongs at each control timescale, what state it can observe, what actions it is authorized to issue, and what evidence is required before those actions reach a live radio system. A non-RT rApp may forecast demand, generate policy, or manage a model over seconds to hours. A near-RT xApp can adjust slicing, power, mobility, or interference decisions on an approximately 10~ms to 1~s loop. Scheduler-local logic and distributed applications (dApps) are proposed for decisions below 10~ms, where transport to a centralized RIC is already too slow~\cite{doro2022dapps}. In this paper, we use the term autonomous network agent to denote an intelligent network function that observes semantically defined network state, reasons over current and predicted conditions, selects actions within explicitly authorized control boundaries, coordinates with other network functions, and adapts or relinquishes authority when its operating assumptions are no longer supported. Such an agent therefore extends beyond a task-specific xApp or rApp: autonomy requires not only decision capability, but also generalization, coordination, assurance, and bounded actuation across control timescales.

Despite this breadth of applications, many existing approaches learn a narrowly defined task from a prescribed set of observations and objectives. A policy trained for network slicing does not automatically provide useful knowledge for mobility management or interference mitigation, even when these functions depend on overlapping network conditions. Changes in traffic, topology, propagation, or available telemetry can also require adaptation within the same task \cite{lotfi2023attention}. Meta-learning, domain-shift-approached and coordinated training, address aspects of this problem, but often retain separate representations and training procedures for individual functions. As AI becomes more widely integrated into the RAN, the cost of developing and maintaining these specialized models raises a broader question: how much knowledge can be shared across network tasks and deployments?

Foundation models offer a potential basis for such reuse through broad pretraining followed by adaptation to downstream tasks. In wireless systems, this could involve learning representations of radio signals, channels, or network behavior that support multiple prediction and control functions \cite{11140266}. Language and multimodal models offer a complementary capability by incorporating service requirements, operational knowledge, and application context into network decisions. Their value, however, must be established through demonstrated transfer, reduced adaptation cost, and improved downstream performance under realistic deployment constraints. 

This paper examines AI-RAN from that perspective, using Open RAN as the architectural setting for deploying and evaluating intelligence. We first organize existing approaches by their prediction, control, coordination, and adaptation roles, relating each role to its observations, actions, and execution timescale. We then examine the limitations of task-specific learning and assess how foundation models could support reusable wireless representations and broader network context. Finally, we identify the data, adaptation, and experimental requirements needed to evaluate this transition. 

\begin{figure}[t]
  \centering
  \includegraphics[width=\columnwidth]{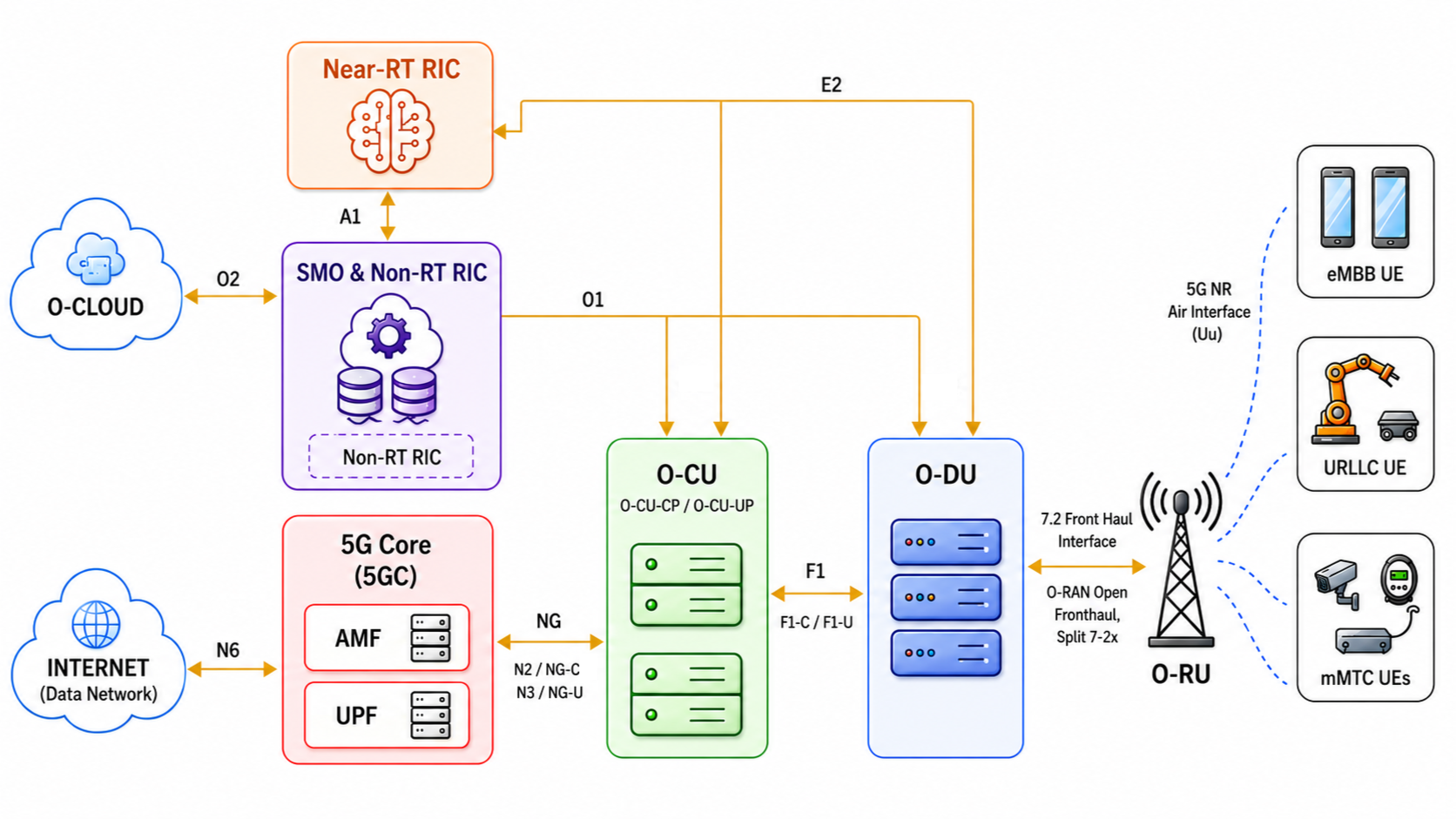}
  \caption{
  Control and learning hierarchy for AI-native Open RAN. Intelligence moves from long-horizon policy and model lifecycle management toward near-RT control, real-time radio logic, and inline execution as deadlines tighten. MEC objectives and assurance evidence must be bound to the same state and action contracts.}
  \label{fig:hierarchy}
  \vspace{-10pt}
\end{figure}

\section{Intelligence Across the O-RAN Control Stack}

\subsection{The O-RAN Architecture}

The O-RAN architecture extends the 3GPP next generation (NG)-RAN through disaggregation, open interfaces, cloud-native deployment, and programmable control. As illustrated in Fig.~\ref{fig:hierarchy}, the RAN is decomposed into the O-RAN Centralized Unit (O-CU), Distributed Unit (O-DU), and Radio Unit (O-RU), while the Service Management and Orchestration (SMO), Non-Real-Time RAN Intelligent Controller (Non-RT RIC), and Near-Real-Time RAN Intelligent Controller (Near-RT RIC) provide management and programmable intelligence~\cite{oran_architecture}.

The O-CU may be divided into control-plane and user-plane functions, with F1 connecting the O-CU and O-DU and NG connecting the NG-RAN to the 5G Core~\cite{3gpp_ts38401}. The O-DU connects to the O-RU through the Open Fronthaul using the 7-2x functional split~\cite{oran_open_fronthaul}. Above these functions, the SMO manages O-RAN functions and cloud infrastructure through O1 and O2~\cite{oran_o1_o2}.
The Non-RT RIC supports long-horizon optimization, policy management, data processing, and AI/ML lifecycle functions. rApps interact with this framework through R1, while A1 conveys policies, enrichment information, and AI/ML guidance toward the Near-RT RIC~\cite{oran_a1_nonrtric}. The Near-RT RIC hosts xApps and interacts with E2 Nodes through E2 Service Models that define available measurements and control operations~\cite{oran_e2_gap}. Together, these components separate data collection, policy coordination, decision making, and RAN actuation across distinct control timescales.

\subsection{Intelligent Control Functions in O-RAN}

O-RAN distributes intelligence across rApps in the Non-RT RIC, xApps in the Near-RT RIC, and emerging dApps colocated with RAN functions. Placement depends on the observations, actions, and latency required by each task. For near-real-time control, E2 Service Model for Key Performance Measurements (E2SM-KPM) provides cell- and UE-level measurements while E2 Service Model for RAN Control (E2SM-RC) exposes supported control services, allowing xApps to observe network state and apply actions supported by the E2 Node.

This framework supports resource allocation and slicing~\cite{kouchaki2022actorcritic,zhang2022teamlearning,zhang2022feddrl,casparsen2025vrslicing}, traffic steering and connection management~\cite{lacava2024trafficsteering,orhan2021gnn}, hierarchical control~\cite{habib2025trafficsteering,habib2023intent}, and energy management~\cite{bordin2025energysaving}. Some tasks require specialized observations beyond standard KPIs. Deployability therefore depends not only on latency but also on whether the required observations and actions are exposed by the RAN.
Faster functions can move closer to the radio. dApps execute alongside O-CU/O-DU functions to shorten the path between observation, inference, and actuation~\cite{doro2022dapps}, with demonstrations in real-time AI control~\cite{lacava2025dapps}, spectrum classification~\cite{olimpieri2025libiq}, and ISAC inference~\cite{polese2026dappisac}. Table~\ref{tab:oran_tasks} summarizes representative mappings among tasks, observations, execution environments, and interfaces.



These mappings distinguish algorithmic performance from deployability. An operational controller must obtain the required telemetry, execute within its deadline, and translate its output into supported RAN actions. Measurements of AI inference inside a Near-RT RIC xApp demonstrate why the complete control loop matters beyond inference latency alone~\cite{obiuwevwi2026realtimeai}. OpenRAN Gym~\cite{bonati2023openrangym}, OAIC~\cite{upadhyaya2023oaic}, and broader testing frameworks~\cite{tang2023aitesting} provide platforms for evaluating these system-level requirements.

\subsection{Closed-Loop Control with Reinforcement Learning}


Reinforcement learning (RL) naturally models closed-loop O-RAN control as repeated interaction between network state and RAN actions. 
Early work applied RL to dynamic resource allocation across O-RAN slices, establishing the basic loop in which observed resource state and service demand determine an allocation whose outcome informs subsequent decisions~\cite{cheng2023oranmidhaul}. More recent work embeds this loop in operational O-RAN systems. REAL integrates an RL-enabled xApp with the O-RAN Software Community (OSC) Near-RT RIC and srsRAN for experimental closed-loop slicing~\cite{barker2025real}. DORA uses online Proximal Policy Optimization (PPO) for dynamic multi-slice resource allocation in an OpenAirInterface-based environment~\cite{dorcheh2025dora}. Other implementations connect DRL resource allocation to executable OpenAirInterface control~\cite{sever2025drlxapp} and adapt near-real-time slicing to channel, traffic, mobility, and QoS conditions~\cite{yan2026xslice}.


Deployability ultimately depends on the complete control loop. Observation intervals determine reaction speed, action spaces determine available control authority, and rewards must balance throughput, latency, fairness, efficiency, and Service Level Agreement (SLA) requirements. Communication and inference delays further create a mismatch between the observed state and the state in which an action is applied. Evaluation should therefore combine learning metrics with end-to-end latency, action frequency, signaling overhead, robustness, and network performance. The progression from slice optimization~\cite{cheng2023oranmidhaul} through experimental and online RL control~\cite{barker2025real,dorcheh2025dora,sever2025drlxapp,yan2026xslice} to integrated traffic-aware control~\cite{sharma2026trafficsteering,groen2026tractor} reflects the broader transition toward continuously operating O-RAN control loops.


\section{Generalization and Reusable Intelligence for Autonomous O-RAN}

RL controllers trained under one set of traffic, channel, mobility, topology, or service conditions may degrade when deployed under another. Generalization concerns preserving useful behavior under such distribution shifts rather than merely adapting within conditions represented during training. Existing O-RAN approaches address this problem through transfer- and meta-learning, robust and distributed learning, environment diversification, transferable representations, and uncertainty-aware adaptation.

\subsection{Transfer Learning and Policy Reuse}


Transfer learning reuses policies, parameters, value estimates, or representations rather than retraining from random initialization. Deep transfer RL has been applied to radio and cache allocation~\cite{zhou2022dtrl}, while O-RAN studies use policy reuse and distillation for slicing~\cite{nagib2024transfer}, repositories of trained agents for new service configurations~\cite{aleyadeh2024transfer}, and policy reuse across non-terrestrial operating conditions~\cite{qazzaz2026xapp}. Its effectiveness depends on similarity between source and target environments, since large shifts can produce negative transfer.

\subsection{Meta-Learning and Rapid Adaptation}

Meta-learning instead trains controllers to adapt rapidly to new tasks. Meta-RL accelerates adaptation across changing O-RAN conditions~\cite{lotfi2025meta}, multi-task initialization supports unseen slicing objectives~\cite{zeng2025m2dqn}, and hierarchical meta-RL combines rapid adaptation with decomposed resource management~\cite{lotfi2025metahrl}. These benefits still depend on training-task diversity, since narrow meta-training may not support changes in topology, mobility, scale, or objectives.

\subsection{Robust Optimization and Safe Generalization}

Robust learning seeks policies that remain effective as conditions change. Sharpness-aware optimization reduces sensitivity to policy perturbations in multi-agent O-RAN resource management~\cite{lotfi2025sam}. SafeSlice~\cite{nagib2025safeslice}, robust SLA-aware control~\cite{yungaicela2026rslaq}, and model-based safe RL~\cite{tuerxun2026saferan} additionally seek to preserve operational constraints during distribution shifts. MORPH broadens training across multiple environments rather than optimizing a Physical Resource Block (PRB) controller for a single network realization~\cite{dorcheh2026morph}. Together, these approaches emphasize that useful generalization requires both performance preservation and safe adaptation.

\subsection{Federated, Multi-Agent, and Population-Based Learning}

Distributed learning broadens the experience available during training. Federated RL shares model knowledge across heterogeneous local conditions without exchanging complete datasets and has been applied to O-RAN slicing and resource allocation~\cite{zhang2022feddrl,mudi2025federated}. Multi-agent approaches decompose resource allocation and coordinated control among interacting policies~\cite{zhang2022teamlearning,habib2023intent,habib2025trafficsteering}. Population-based methods further diversify policies through federated neuroevolution~\cite{kouchaki2025neuroevolution} and evolutionary RL~\cite{lotfi2022evolutionary}. Their generalization benefit ultimately depends on whether training exposes the policies to meaningfully different conditions.

\subsection{Digital Twins and Environment Diversification}

Digital twins and programmable testbeds expose controllers to broader variations in traffic, propagation, mobility, topology, interference, and resources without unsafe exploration on production infrastructure. Proposed RAN twins~\cite{vila2023digitaltwin}, Colosseum~\cite{polese2024digitaltwin}, and OpenRAN Gym~\cite{bonati2023openrangym} support repeatable training and validation. Reliable transfer nevertheless requires both environment diversity and sufficient fidelity to deployment conditions.


\subsection{Uncertainty-Aware and Model-based Adaptation}

Generalizable controllers should also recognize unfamiliar conditions. Bayesian RL incorporates uncertainty into action selection and has been applied to joint O-RAN and Multi-access Edge Computing (MEC) orchestration~\cite{murti2024bayesian}. Model-based RL complements this capability by learning network dynamics for planning and safer adaptation, including model-based safe slicing designed to reduce SLA violations~\cite{tuerxun2026saferan}. Uncertainty and learned dynamics can therefore trigger conservative behavior, adaptation, policy replacement, or higher-level intervention when deployment conditions depart from training experience.

\subsection{Representation-based Generalization}

Generalization also depends on how network state is represented. Attention-based slicing emphasizes relevant portions of network state~\cite{lotfi2023attention}, while predictive methods incorporate learned traffic dynamics into RL decisions~\cite{lotfi2023lstm}. More recent approaches introduce semantic context. LLM-augmented DRL incorporates contextual information~\cite{lotfi2025llm}, prompt tuning adapts this representation to resource management~\cite{lotfi2025prompt}, and ORAN-GUIDE combines retrieval, learned prompting, and RL~\cite{lotfi2025guide}. Wireless foundation models extend this direction through reusable features across wireless tasks~\cite{11140266}, while ORANSight-2.0 explores foundation models for O-RAN knowledge and reasoning~\cite{gajjar2025oransight}.

These approaches shift generalization from task-specific policies tied to observations toward representations that capture relationships among radio conditions, traffic, resources, service requirements, and network context. The next challenge is whether such representations can support multiple control functions and ultimately provide a common abstraction not only for observing the RAN but also for acting upon it.


\subsection{Foundation Models, World Models, and Reusable Network Intelligence}

Reusable representations alone do not provide general control. Most RL controllers still operate within predefined state and action spaces, while transfer and meta-learning generally assume compatible tasks or policies. The remaining challenge is to connect representations learned across environments and network functions to controllers with different observations, objectives, and actuation capabilities.

The emergence of wireless foundation models (WFMs) marks a paradigm shift from fragmented, task-specific deep learning toward unified, adaptable architectures for next-generation (6G) networks. Pre-trained on large, unlabeled radio frequency (RF) datasets spanning raw in-phase/quadrature (IQ) samples, channel state information (CSI), and spectrograms using self-supervised learning \cite{alikhani2025lwm, zhou2025spectrumfm}, these architectures extract universal physical-layer representations. Adopting foundation models in telecommunications bridges long-standing operational silos, unifying core physical-layer operations like channel estimation, beam prediction, and signal identification with integrated sensing and communication (ISAC) and agentic network orchestration.

This adoption is of conventional supervised models that require expensive data labeling and suffer acute performance degradation in dynamic, out-of-distribution wireless environments. By contrast, WFM decouples universal feature extraction from downstream task execution. This enables powerful zero-shot and few-shot transfer learning, allowing networks to adapt to unseen propagation environments or emerging tasks with minimal parameter fine-tuning \cite{uddin2026rfprompt}. Ultimately, adopting foundation models provides the computational efficiency, generalizability, and multi-task scalability required to realize truly autonomous, AI-native wireless ecosystems.




Transferable representations provide reusable state abstractions, but they do not by themselves capture how the network evolves under candidate control actions. Model-free policies similarly select actions without explicitly predicting their consequences. World models complement these approaches by learning network dynamics, allowing a decision to be evaluated in imagination before execution \cite{ha2018world} and recent works learn latent dynamics over KPI and resource trajectories to support planning without online trials~\cite{rezazadeh2025agentic,zhou2025slicewm}. 


\begin{table*}[t]
\caption{Representative Mapping of Intelligent RAN Functions to O-RAN Applications}
\label{tab:oran_tasks}
\centering
\scriptsize

\setlength{\tabcolsep}{3pt}
\renewcommand{\arraystretch}{0.95}

\begin{tabularx}{\textwidth}{
    @{}
    p{0.13\textwidth}
    p{0.20\textwidth}
    p{0.105\textwidth}
    p{0.17\textwidth}
    Y
    @{}
}
\toprule
\textbf{RAN function} &
\textbf{Representative observations} &
\textbf{Application} &
\textbf{Interface/service requirement} &
\textbf{Representative work} \\
\midrule

Resource allocation &
Traffic, CQI, throughput, resource utilization &
xApp &
E2 telemetry and RAN control &
Actor--critic~\cite{kouchaki2022actorcritic}, team learning~\cite{zhang2022teamlearning} \\

Network slicing &
Slice traffic, latency, throughput, PRB utilization &
xApp/rApp &
E2 control and A1 policy &
Federated DRL~\cite{zhang2022feddrl}, ORANSlice~\cite{cheng2024oranslice}, VR slicing~\cite{casparsen2025vrslicing} \\

Traffic steering &
UE radio measurements, throughput, cell load &
xApp &
E2SM-KPM and E2SM-RC &
Traffic steering~\cite{lacava2024trafficsteering}, hierarchical control~\cite{habib2025trafficsteering} \\

Connection management &
UE--cell measurements, load, interference &
xApp &
E2 monitoring and control &
GNN/RL xApp~\cite{orhan2021gnn} \\

Intent-driven orchestration &
Service intent and network state &
rApp/xApp &
Non-RT policy and near-RT control &
Hierarchical orchestration~\cite{habib2023intent} \\

Energy management &
Load, traffic history, utilization &
rApp/xApp &
Policy and near-RT control &
DRL energy saving~\cite{bordin2025energysaving}, ScalO-RAN~\cite{maxenti2024scaloran} \\

Spectrum sensing &
Spectrum/PHY observations, interference state &
xApp/dApp &
E2 or lower-layer data exposure &
SenseORAN~\cite{reusmuns2024senseoran}, LibIQ~\cite{olimpieri2025libiq} \\

Localization &
SRS channel estimates, radio measurements &
xApp &
SRS-oriented E2 service model &
OAI/FlexRIC localization~\cite{bouknana2026localization} \\

Fast PHY/MAC control &
PHY state, I/Q, channel information &
dApp &
Low-latency RAN interface &
dApps~\cite{doro2022dapps,lacava2025dapps}, ISAC inference~\cite{polese2026dappisac,villa2026isac} \\

\bottomrule
\end{tabularx}
\end{table*}

\section{Multi-Access Edge Computing}
\label{sec:mec}

The transition toward autonomous O-RAN intelligence extends control beyond radio resources to the applications they serve. MEC introduces service placement, application execution, and computing contention into this decision process. Low packet latency alone does not ensure timely application completion: transmission, queueing, and execution consume a shared deadline budget \cite{xu2022tutti,yi2025arma,zhang2026smec}. We therefore organize MEC research along three complementary problem dimensions-service placement and continuity, task offloading and application adaptation, and resource allocation and execution-and then consider cross-layer coordination and end-to-end service assurance as a synthesis dimension. As in Sections II and III, the relevant distinctions are the information available to each controller, its actuation scope, and its operating timescale.

\subsection{Architecture, Placement, and Service Continuity}

The MEC architecture separates hosts, comprising the platform and virtualized infrastructure, from system-level orchestration responsible for host selection and application lifecycle management \cite{etsi2024mec003}. In 5G, local user plane routing through an appropriately selected User Plane Function (UPF) provides connectivity to edge application servers, complemented by application-server discovery and traffic-steering or relocation procedures \cite{3gpp2025ts23548}. These mechanisms extend the service environment around the RIC architecture in Section II. MEC application management, 5G session control, and RAN control retain distinct responsibilities even when their functions share infrastructure.

Service placement determines where application instances and their state reside; resource provisioning determines the capacity available to execute their workloads. Joint orchestration formulations connect hosting locations, O-RAN functional splits, routing, and computing allocations to infrastructure cost and delay \cite{murti2024bayesian}. Placement must also accommodate demand variation: robust formulations distinguish advance resource reservation from subsequent placement and workload adaptation, accounting for spatial and temporal demand correlations \cite{cheng2025robust}. This introduces a planning problem that extends beyond the instantaneous radio state.


\subsection{Offloading, Adaptation, and Heterogeneous Execution}

Task offloading connects application demand to communication and computing budgets. Learning-based policies can jointly decide whether tasks execute locally or are offloaded to a MEC server, while allocating communication and computational resources under differentiated latency and energy objectives \cite{ebra2025offloading}. For AI services, this action space extends to the workload itself. Adapting video bitrate and inference model complexity changes transmission and execution demand while preserving an application-specific accuracy requirement \cite{yi2025arma}. Fine-grained DNN partitioning further distributes execution across devices and edge servers, making intermediate-feature transmission and heterogeneous computing capabilities part of the latency trade-off \cite{li2024dnnpartition}. These decisions couple where an application runs with how it is executed.

Allocated computing capacity, however, does not directly imply predictable service time. On shared accelerators, GPU operator characteristics and interference between concurrent workloads affect execution latency, motivating fine-grained scheduling rather than a uniform computing-cycle abstraction \cite{strati2024orion}. Deadline-sensitive MEC additionally requires visibility into request progress and mechanisms to adjust CPU allocation and GPU priorities at runtime \cite{zhang2026smec}. The generalization problem from Section III consequently extends to changes in model configuration, execution placement, and hardware contention as well as wireless channels. A transferable policy must retain an adequate representation of these changing constraints.

\subsection{Cross-Layer Coordination and Service Assurance}

Application-aware scheduling provides a bridge between service objectives and RAN actuation. Application context and channel estimates can be combined to anticipate frame demand and prioritize radio grants according to approaching deadlines \cite{xu2022tutti}. Extending this approach across the application and network allows resource information and latency budgets to inform both radio and GPU scheduling \cite{yi2025arma}. Such coordination exposes dependencies that are obscured when transmission and execution are optimized independently.

Explicit cross-layer exchange is one coordination approach; local inference is another. Request activity can be inferred at the RAN from standard 5G control signals, while probing and application lifecycle instrumentation provide the edge with estimates of the remaining deadline budget. Independent schedulers can then adapt radio resources, CPU allocation, GPU priorities, and early dropping without direct RAN--edge coordination \cite{zhang2026smec}. The resulting architectural choice concerns how much state is exchanged, how accurately unobserved progress can be inferred, and whether information remains useful when an action takes effect. Neither explicit coordination nor local inference removes the need to account for observation error and control delay.


\subsection{Implications for Autonomous O-RAN Agents}

Hierarchical coordination has already been investigated in O-RAN: operator objectives can guide the selection of lower-level xApps through hierarchical reinforcement learning \cite{habib2023hierarchical}, while proposed dApp architectures extend control to RAN-local execution under tighter timing constraints \cite{lacava2025dapps}. Recent agentic frameworks further place intent interpretation and model governance in longer-timescale control loops, coupled with faster control and inference functions \cite{navidan2026agentic}. These precedents provide an architectural basis for discussing MEC integration; the hierarchy itself should not be regarded as a new contribution of this roadmap.

For MEC, the additional requirement is to connect these control layers to service placement and application execution. Joint O-RAN/MEC orchestration and adaptive placement formulations establish relevant provisioning decisions \cite{murti2024bayesian,cheng2025robust}, while learning-based task-offloading and resource allocation policies \cite{ebra2025offloading} and application-aware MEC systems expose complementary mechanisms for coordinating radio and execution deadlines \cite{yi2025arma,zhang2026smec}. Building on this evidence, we identify the coordination of placement, application configuration, and runtime scheduling as a design requirement for autonomous RAN--MEC operation. Application lifecycle information and accelerator controls must be integrated through the relevant MEC and runtime mechanisms \cite{zhang2026smec,etsi2024mec003}, rather than inferred to be available merely because a RAN control interface exists.


The research question is therefore how to compose these capabilities under an end-to-end service objective. For the roadmap considered here, we propose evaluating this composition under workload shifts, mobility, stale observations, and resource failures, with explicit checks on control authority and defined handling of infeasible actions. These are evaluation requirements of our synthesis, not guarantees established by the cited systems. They connect MEC service assurance to the trust and control-security questions addressed in Section~\ref{sec:zero-trust}.

\section{Zero Trust Security in Open RAN}
\label{sec:zero-trust}


The same interoperable interfaces that enable ORAN programmability also widen its threat surface~\cite{zt_barker2026survey}. A1 policy, E2 telemetry and control, O1/O2 management, and third-party xApps introduce trust boundaries alongside inherited 5G New Radio vulnerabilities~\cite{polese2023understanding,zt_oran_security2026}. SNI5GECT demonstrates pre-authentication injection, software attacks can compromise insufficiently protected packetized fronthaul, and sidelink and location-aware attacks expose additional radio-local threats~\cite{zt_luo2025sni5gect,zt_xing2024fronthaul,zt_erni2025glados}. Such attacks can corrupt control inputs or selectively deny service without compromising the 5G Core, making deadline and service failures relevant security outcomes.

Zero Trust couples continuous verification with least-privilege authorization~\cite{9162761}. Interface protections provide a transport baseline, while OZTrust and XRF restrict xApp communication and service access~\cite{zt_oran_security2026,zt_groen2024interfaces,zt_jiang2023oztrust,zt_atalay2023xrf}. Yet authenticated actors and protected transport do not establish that subsequent telemetry is semantically truthful or bound to the current UE and bearer context~\cite{zt_barker2026survey}.



This distinction becomes critical for learning-enabled control. A backdoored deep reinforcement learning (DRL) xApp can behave normally until poisoned observations activate harmful actions~\cite{zt_lacava2025poison}. ORAN-DEFEND applies subspace-based sanitization to frozen black-box O-RAN policies by using trusted trajectories to identify a safe observation subspace, projecting incoming KPI windows toward that subspace before inference, and using the projection residual as an anomaly signal~\cite{zt_bharti2022defense,uddin2026defend}. Evaluation on Colosseum ColO-RAN telemetry shows that such sanitization can substantially recover policy behavior for detectable triggers, while performance degrades when malicious perturbations lie within the learned safe subspace~\cite{polese2023coloran,uddin2026defend}. This limitation illustrates a broader principle for autonomous O-RAN: anomaly mitigation should inform and bound control authority rather than establish unconditional trust.

Receiver feedback forms an important trust boundary because reports such as the Channel Quality Indicator (CQI), Rank Indicator (RI), Precoding Matrix Indicator (PMI), and Hybrid Automatic Repeat reQuest (HARQ) directly influence scheduling and link adaptation~\cite{zt_3gpp38214}. Prior work shows that channel information can be cross-checked using acknowledgement/negative-acknowledgement (ACK/NACK) feedback~\cite{zt_wiesmayr2026salad}, while deceptive RI signaling can distort Proportional Fair (PF) scheduler behavior and disadvantage honest users through its effect on scheduler memory~\cite{zt_timilsina2025csi,zt_barker2026starvation}. These results show that authenticated or protocol-valid feedback should not automatically be treated as trustworthy state. For autonomous O-RAN control, independently observable evidence should therefore be used to assess reported state and to restrict the authority of a policy when the evidence is inconsistent.

These results establish a progression from authenticating actors to validating evidence and finally bounding control authority. Lifecycle governance and cross-site analysis can remain in non-RT control, reversible adaptation can operate in the near-RT layer, and only deadline-critical checks should move toward local execution~\cite{zt_barker2026survey,doro2022dapps}. Deployment therefore requires fresh evidence, explicit UE and bearer scope, bounded actions, and tested fallback behavior. Security evaluation must likewise account for observation age and complete observation-to-actuation delay rather than inference latency alone~\cite{barker2026atlasran}.

\section{Discussion and Open Research Questions}
The reviewed literature suggests that progress toward autonomous Open RAN depends not only on increasingly capable models, but on whether intelligence can generalize across operating conditions, compose across control functions and timescales, act within evidence-supported authority, and satisfy measurable deployment constraints. From this perspective, five research challenges define the transition from task-specific xApps and rApps toward autonomous network agents.

The first is \emph{generalization with bounded authority}. Transfer learning, meta-learning, robust optimization, multi-environment training, and foundation model representations reduce adaptation cost, but none eliminates distribution shift. Future evaluations should therefore characterize a transfer envelope over traffic, channel conditions, topology, software configuration, available telemetry, and execution environment. Controllers should detect operation outside this envelope and reduce their action authority or invoke a validated fallback rather than extrapolate with unchecked confidence.

The second challenge is \emph{composition across intelligent functions}. Multiple rApps, xApps, dApps, and MEC agents may produce individually reasonable decisions that compete for the same radio or computing resources. Learning-based coordination alone does not resolve this runtime problem. Resources such as PRBs, power, handover parameters, accelerator capacity, and model placement require explicit scope, priority, and validity intervals. Arbitration must also distinguish conflicting policies from malicious or untrusted inputs because both can produce unsafe actions but require different recovery mechanisms. Authentication should therefore establish origin without implying semantic validity or unrestricted control authority~\cite{zt_barker2026survey}.

The third challenge is evidence-conditioned control authority. Increasing model capability does not imply that a controller should exercise unrestricted authority. An autonomous agent may operate with stale observations, unfamiliar network conditions, conflicting policies, unavailable actuators, or evidence whose integrity cannot be independently established. Future systems therefore need mechanisms that translate the quality and freshness of available evidence into explicit control boundaries: which actions are permitted, over which resources and users, for how long, and with what fallback when confidence degrades. This shifts Zero Trust from authentication alone toward a runtime relationship between evidence, confidence, and actuation authority.

The fourth challenge is \emph{experimentally credible autonomy}. Simulation, emulation, software-defined-radio systems, and live O-RAN platforms preserve different aspects of network behavior and should support correspondingly different claims. Evaluation should separate four relevant clocks. The radio clock governs scheduling and retransmission, the control clock spans observation through actuation, the application clock determines whether completed service retains value, and the learning clock governs model adaptation and promotion. Reporting inference latency alone can therefore make a controller appear deployable even when its observations are stale or its actions miss radio or application deadlines. Relevant measurements include state age, end-to-end control latency, actuation success, deadline success, radio and compute utilization, and recovery under failed or untrusted inputs~\cite{barker2026atlasran}.

The fifth challenge is \emph{semantic interoperability}. Portable models are insufficient when implementations disagree about the meaning of observations and actions. A deployable controller should carry a machine-readable contract describing input semantics, normalization, update rate, action units, target scope, dependencies, validated operating envelope, and fallback behavior. Model and policy versions should remain bound to the service-model schema and software stack on which they were validated. O-RAN interfaces provide communication interoperability, but autonomous operation additionally requires preservation of these state and action semantics across O1, A1, E2, local APIs, and application services~\cite{polese2023understanding}.

These requirements also define a practical path toward increasing autonomy. Controllers can progress from offline evaluation to shadow operation and then to bounded control over selected cells, slices, or services. Promotion should depend on stored evidence and support rollback, while execution should produce an audit trail connecting observed state, selected policy, issued action, outcome, and fallback status. Security evaluation should similarly distinguish algorithmic recovery from service recovery because success in KPI replay or slot-level simulation does not establish end-to-end containment under live timing constraints~\cite{uddin2026defend,zt_barker2026starvation,barker2026atlasran}.

Foundation models fit naturally into this architecture, but their near-term role is more compelling in representation, planning, and coordination than in direct fast-timescale actuation. They can translate human intent and network knowledge into policies, retrieve operational context, select or configure specialist models, and provide reusable RF representations~\cite{gajjar2025oransight,lotfi2025prompt,lotfi2025guide,uddin2026rfprompt}. Near-real-time and lower-layer actions should remain grounded through typed interfaces and bounded controllers whose timing and behavior can be validated. The roadmap toward autonomous O-RAN is therefore not a progression from small models to larger models, nor from xApps to a single monolithic agent. It is a progression from isolated task-specific controllers toward transferable and predictive intelligence that can coordinate specialized functions across timescales while acquiring, exercising, and relinquishing control authority according to current evidence and validated system constraints.


\section{Conclusion}

AI-native Open RAN is evolving from isolated optimization applications toward coordinated network agents operating across multiple control timescales. The reviewed literature traces this progression through adaptive and generalizable RL, semantic representations, closed-loop xApps, MEC and radio-compute orchestration, accelerated execution, and security assurance. Across these developments, increasing autonomy requires more than increasingly capable models. It requires transferable state representations, explicit action semantics, measurable timing, bounded authority, and validated fallback behavior. rApps, xApps, dApps, MEC agents, and inline functions can therefore operate as complementary components of a common control architecture rather than independent models competing for resources and authority. Foundation models can broaden representation, reasoning, and coordination, while time-critical actuation remains grounded in bounded and verifiable controllers. Progress toward autonomous Open RAN will ultimately depend on whether intelligence can generalize across deployments while preserving interoperability, timing guarantees, and control authority supported by current evidence.

\bibliographystyle{IEEEtran}
\bibliography{references}

\end{document}